\documentclass[letterpaper,usenatbib]{mn2e}
\usepackage{amsmath}
\usepackage{epsfig}
\usepackage{multirow}
\usepackage{times}
\usepackage{verbatim}
\def\gtsim{~\rlap{$>$}{\lower 1.0ex\hbox{$\sim$}}}
\def\ltsim{~\rlap{$<$}{\lower 1.0ex\hbox{$\sim$}}}

\title[ALMA observations of NGC 253]{ALMA observations of 99 GHz free-free and H40$\alpha$ line emission from star formation in the centre of NGC~253}

\author[G. J. Bendo et al.]
    {G. J. Bendo$^{1,2}$, R. J. Beswick$^{1,2}$, M. J. D'Cruze$^1$, C. Dickinson$^1$, G. A. Fuller$^{1,2}$,\newauthor T. W. B. Muxlow$^{1,2}$\\ 
    $^1$   Jodrell Bank Centre for Astrophysics,
           School of Physics and Astronomy, The University of Manchester, 
           Oxford Road, Manchester M13 9PL, UK\\
    $^2$   UK ALMA Regional Centre Node
}
\date{}
\pagerange{\pageref{firstpage}--\pageref{lastpage}}
\pubyear{}

\begin{document}
\label{firstpage}
\maketitle

\begin{abstract}
We present Atacama Large Millimeter/submillimeter Array observations of 99.02 GHz free-free and H40$\alpha$ emission from the centre of the nearby starburst galaxy NGC~253.  We calculate electron temperatures of 3700-4500~K for the photoionized gas, which agrees with previous measurements.  We measure a photoionizing photon production rate of$(3.2\pm0.2)\times10^{53}$~s$^{-1}$ and a star formation rate of $1.73\pm0.12$~M$_\odot$~yr$^{-1}$ within the central 20$\times$10~arcsec, which fall within the broad range of measurements from previous millimetre and radio observations but which are better constrained.  We also demonstrate that the dust opacities are $\sim$3 dex higher than inferred from previous near-infrared data, which illustrates the benefits of using millimetre star formation tracers in very dusty sources.
\end{abstract}


\section{Introduction}
\label{s_intro}
\addtocounter{footnote}{3}

\begin{figure*}
\epsfig{file=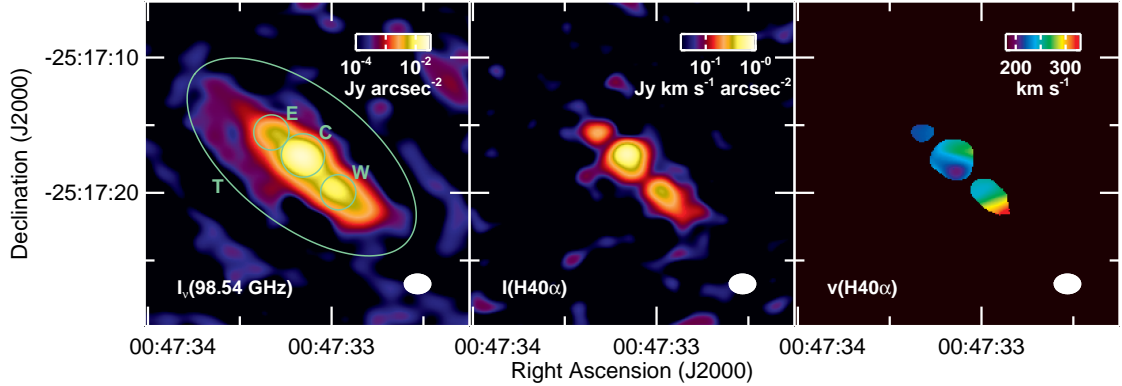}
\caption{Images of the continuum surface brightness, the H40$\alpha$ intensity, and the H40$\alpha$ mean velocity in the central 32$\times$32~arcsec of NGC~253.  The velocity, which is relative to the Solar System Barycentre, is only shown for data where the H40$\alpha$ intensity is detected at the $5\sigma$ level.  The while oval at the bottom right of each panel shows the FWHM of the beam.  The green regions in the continuum image show the total (T), east (E), central (C), and west (W) regions within which fluxes and spectra were measured.  The spectra are shown in Figure~\ref{f_spec}, and measured quantities are listed in Table~\ref{t_spec}.}
\label{f_map}
\end{figure*}

The millimetre waveband contains two star formation tracers that directly trace photoionizing light from star forming regions while not suffering from dust obscuration effects, making it a superior alternative to star formation tracers in other wavebands (see \citet{calzetti09} and \citet{murphy11} for reviews).  The continuum emission includes free-free emission, and while thermal dust emission is dominant at $>$100~GHz and synchrotron emission is more prominent at $\leq$30~GHz, free-free emission is the dominant emission source at 30-100~GHz \citep[e.g][]{peel11}.  Multiple hydrogen recombination lines are seen at millimetre wavelengths as well.  Optical and near-infrared lines are affected by dust extinction effects, and centimetre and metre recombination lines are affected by potential masing and collisional broadening effects, but millimetre lines are not affected by any of these problems \citep[e.g.][]{gordon90} and are therefore more accurate tracers of star formation.  

The Atacama Large Millimeter/submillimeter Array (ALMA) is a groundbreaking millimetre and submillimetre telescope with the capability of reaching sensitivity levels in the 30-950 GHz range at least an order of magnitude better than other telescopes (see \citealt{lundgren13} for a technical overview).  ALMA is capable of detect free-free and recombination line emission from many nearby infrared-luminous galaxies \citep{scoville13}.  It is therefore of interest to use early ALMA observations to explore the telescope's capabilities to detect these star formation tracers.

In this paper, we present ALMA observations of free-free continuum emission and H40$\alpha$ emission at 99.02~GHz from the centre of NGC~253 that we use to derive  star formation rates (SFRs) as well as the electron temperature ($T_e$) of the gas.  In addition, we use the H40$\alpha$ data together with near-infrared data from \citet{engelbracht98} to examine the dust attenuation.  \citet{bolatto13}, \citet{leroy15}, and \citet{meier15} previously published analyses based on these data, but except for a brief mention of recombination lines in the latter paper, they focused on the molecular gas, whereas we will concentrate on the photoionized gas.  

Radio and millimetre recombination lines have previously been detected from NGC~253 \citep[e.g.][]{seaquist77, mebold80, anantharamaiah96, puxley97, mohan02, mohan05, rodriguezrico06, kepley11}.  This includes the detection of H40$\alpha$ emission by \citet{puxley97}.  However, our analysis will focus on millimetre lines that are less susceptible to masing effects than most previous recombination line observations, which have been at lower frequencies.  Additionally, most of the continuum at the location of the H40$\alpha$ line is free-free emission, allowing us to use the line and continuum data together to measure $T_e$, whereas the lower frequency data are more affected by synchrotron emission.  The ALMA data also have superior spatial resolutions, spectral resolutions, and sensitivities than most previously-published NGC~253 data.  This ultimately serves as an example of the future work on extragalactic star formation in nearby galaxies that can be done with ALMA.

\section{Data}

The data were originally acquired as part of program 2011.0.00172.S and published by \citep{bolatto13}.  The observations covering the H40$\alpha$ line are from three execution blocks (EBs) performed on 07 May, 01 July, and 02 July 2012.  Three locations along the major axis spaced by 25~arcsec were observed, and the total integration time per location was 29 min.  Only 14-20 antennas were operational.  The spectral window covering the H40$\alpha$ line was centered at a sky frequency of 98.54~GHz, contained 3840 channels each with a width of 488 kHz (1.5 km~s$^{-1}$), and included both polarisations.  Uranus was the flux calibrator, J2333-237 was the bandpass calibrator, and J0137-245 was the phase calibrator.  

We reprocessed the data using the Common Astronomy Software Applications version 4.2.2, which includes a version of the flux calibration (Butler-JPL-Horizons 2012) that is more up-to-date than the one used to create the data in the ALMA archive.  The visibility data were processed through steps to flag bad data and to calibrate the phase and amplitude as a function of frequency and time.  Next, we rescaled the amplitudes of the visibility data so that the signal measured for J0137-245 in the individual execution blocks matched the weighted average of these values, and then we concatenated the data.  We produced two versions of the image cube (with and without continuum) using a clean algorithm with natural weightings in an interactive mode.  Although the primary beam is 63~arcsec, we created image cubes covering a 32$\times$32~arcsec region (with 0.1~arcsec pixels) because no detectable emission is found outside this region.  The cubes used in the analysis cover a sky frequency range between 98.80 and 99.09~GHz, and each channel in each cube covers 8 channels in the visibility data, which corresponds to 3.9~MHz ($\sim$11.8~km~s$^{-1}$).  For display purposes, we also produced a cube that covered 97.64-99.44~GHz, which was the full usable range of the spectral window containing the H40$\alpha$ line, with the same spatial and spectral resolution.  The full-width at half maximum (FWHM) of the reconstructed beam is $1.9\times1.6$~arcsec; assuming a distance of $3.44\pm0.26$~Mpc (based on the average of distances measured by \citealt{dalcanton09}), this corresponds to spatial scales of $\sim30$~pc.  The flux calibration uncertainty reported in the ALMA Technical Handbook is 5\% \citep{lundgren13}.  The 98.54~GHz flux densities for the bandpass calibrator ($1.15\pm0.04$~Jy) and the phase calibrator ($1.09\pm0.06$~Jy) have relative uncertainties (based on the variations in measurements from individual EBs) that are consistent with the Technical Handbook, although unidentified systematic amplitude calibration effects could affect the flux densities.  Cleaned image cubes of the bandpass and phase calibrators show $<$1\% variations in the flux density from channel to channel, and no spectral features are seen at the position of the H40$\alpha$ line in the visibility data for the calibration sources, which indicates that any line emission $>$1\% of the continuum emission is not caused by calibration issues.

We had access to data from program 2011.0.00061.S (PI: Takano) that also covers the H40$\alpha$ line.  Because the spectral settings were different from program 2011.0.00172.S, we did not use the data in our analysis, but we did examine the processed archival data for program  2011.0.00061.S to confirm that the H40$\alpha$ emission is detected in those data.

\begin{figure}
\epsfig{file=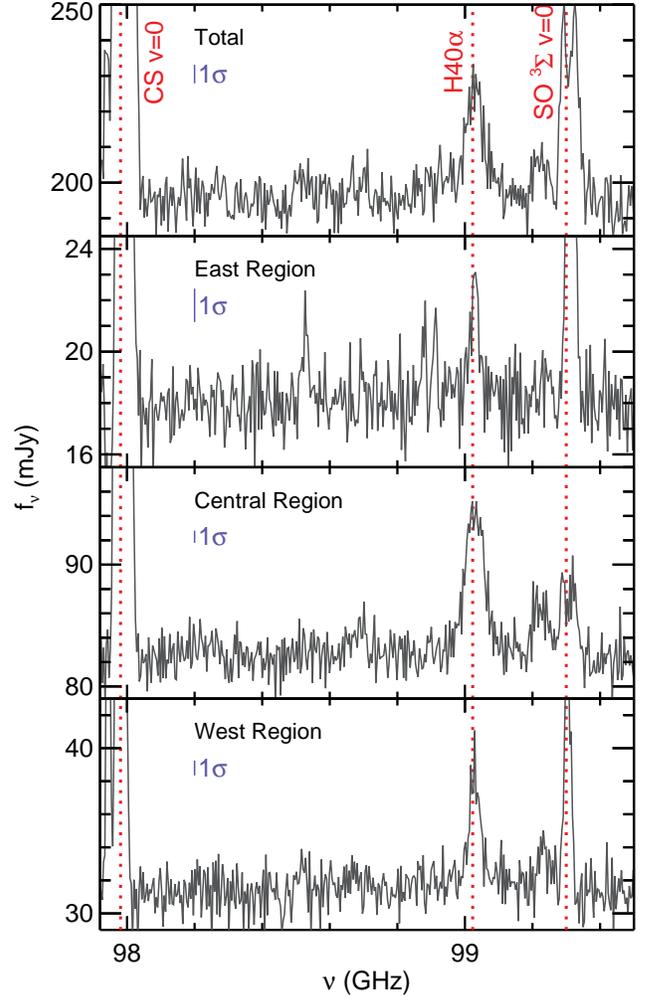}
\caption[]{Plots of the spectra measured in the 20$\times$10~arcsec elliptical region (labelled as ``Total'') and in the east, central, and west subregions identified in Figure~\ref{f_map}.  The frequencies are rest frequencies based on the velocities in Table~\ref{t_spec}.  The red lines show major spectral features identified using {\sc Splatalogue}\footnotemark[3].  The blue bar shows the noise per channel as listed in Table~\ref{t_spec}.}
\label{f_spec}
\end{figure}

H40$\alpha$ fluxes, mean velocities (relative to the Solar System Barycentre, and velocity FWHM were measured in the continuum-subtracted image cube, and the continuum at 99.02~GHz was measured in the other image cube by fitting and removing the line emission.  Images of the the continuum emission (based on all continuum data within the spectral window), the H40$\alpha$ flux, and the H40$\alpha$ mean velocity are shown in Figure~\ref{f_map}.  Spectra integrated over the central $20\times10$~arcsec as well as integrated within regions covering the three brightest regions are shown in Figure~\ref{f_spec}.  Details of the measurements in these regions are listed in Table~\ref{t_spec}.  The uncertainties in the measurements incorporate the noise per channel values listed in Table~\ref{t_spec}, but except for the east (E) region, the accuracy in the continuum and H40$\alpha$ fluxes is primarily limited by the calibration uncertainties.  Most of the signal outside the $20\times10$~arcsec ellipse is detected at $\ltsim$3$\sigma$, and we lack the {\it uv} coverage to recover $>$20~arcsec structures.

The central star forming region can be divided into three knots that lie along the major axis of the galaxy.  The central (C) and west (W) sources are detected at $>$10$\sigma$ in both the continuum and H40$\alpha$ emission.  The eastern source is detected at $>$10$\sigma$ in continuum emission as well, but the peak is only detected at the $\sim$8$\sigma$ level in H40$\alpha$ emission.  Although all three sources had previous been detected in radio continuum emission \citep{mohan05, kepley11}, only \citep{anantharamaiah96} had previously shown radio recombination line emission from all three sources.  The same structure is also seen in the Br$\gamma$ image from \citet{engelbracht98}, but the source is only marginally resolved and does not appear to show the same peak in recombination line emission.  

The H40$\alpha$ emission from the central source is relatively broad (with a FWHM of $191\pm4$~km s$^{-1}$) and appears to be rotating orthogonally to the plane of the galaxy, which is consistent with previous radio recombination line observations \citep{anantharamaiah96, rodriguezrico06, kepley11}.  This may indicate that material has recently fallen into the central region, that the central source comprises two superimposed photoionized regions, or that the ionized gas is in a superwind that is flowing in a direction aligned with the minor axis.  It is unlikely that an active galactic nucleus is present, as the H40$\alpha$ line would appear $\gtsim$1000~km~s $^{-1}$ in width.    We do not detect any asymmetry in the line emission from the central source as had been reported in some prior recombination line observations \citep{mohan02, kepley11} but not others.  The reason for this discrepancy is unclear, although one possibility is that the line shapes differ because the lower-frequency lines are probing less dense gas than the H40$\alpha$ line.

\section{Analysis}

\begin{table*}
\centering
\begin{minipage}{167mm}
\caption{Measurements of the continuum and H40$\alpha$ emission.}
\label{t_spec}
\begin{tabular}{@{}lccccccccc@{}}
\hline
Region &
    Noise &
    99.02~GHz &
    H40$\alpha$ &
    H40$\alpha$ &
    H40$\alpha$ &
    H40$\alpha$ &
    Electron &
    \multicolumn{2}{c}{SFR} \\
&
    per &
    Free-Free Flux &
    Flux &
    Mean &
    FWHM &
    /Free-Free &
    Temperature$^b$ &
    Free-Free$^b$ &
    H40$\alpha$ \\
&
    Channel$^a$ &
    Density$^b$ &
    &
    Velocity &
    &
    Ratio$^b$ &
    &
    &
    \\
&
    (mJy) &
    (mJy) &
    (Jy km s$^{-1}$)&
    (km s$^{-1}$) &
    (km s$^{-1}$) &
    (km s$^{-1}$) &
    (K) &
    (M$_\odot$ yr$^{-1}$) &
    (M$_\odot$ yr$^{-1}$) \\
\hline
Total &
  5.2 &
  $146.5 \pm 17.2$ &
  $6.91 \pm 0.80$ &
  $244 \pm 2$ &
  $169 \pm 10$ &
  $47.2 \pm 2.4$ &
  $3900 \pm 300$ &
  $1.59 \pm 0.16$ &
  $1.87 \pm 0.18$ \\
East &
  1.3 &
  $13.4 \pm 1.5$ &
  $0.64 \pm 0.08$ &
  $217 \pm 4$ &
  $85 \pm 11$ &
  $48.0 \pm 3.3$ &
  $3900 \pm 400$ &
  $0.15 \pm 0.02$ &
  $0.17 \pm 0.02$ \\
Central &
  1.0 &
  $58.6 \pm 7.0$ &
  $2.49 \pm 0.28$ &
  $236 \pm 2$ &
  $191 \pm 5$ &
  $42.5 \pm 2.0$ &
  $4500 \pm 300$ &
  $0.58 \pm 0.05$ &
  $0.63 \pm 0.06$ \\
West &
  0.9 &
  $22.8 \pm 2.7$ &
  $1.13 \pm 0.15$ &
  $258 \pm 1$ &
  $113 \pm 8$ &
  $49.5 \pm 4.1$ &
  $3700 \pm 400$ &
  $0.25 \pm 0.03$ &
  $0.29 \pm 0.02$ \\
\hline
\end{tabular}
$^a$ The root mean square noise levels were measured on either side of the H40$\alpha$ line at rest frequencies (based on v=236 km s$^{-1}$) of 98.880-98.920 and 99.135-99.175~GHz (excluding the unideintified spectral feature at 98.915~GHz in the east region).\\
$^b$ These numbers are based on multiplying the measured continuum by $0.70\pm0.10$ to account for emission from sources other than free-free emission.
\end{minipage}
\end{table*}

\begin{figure}
\epsfig{file=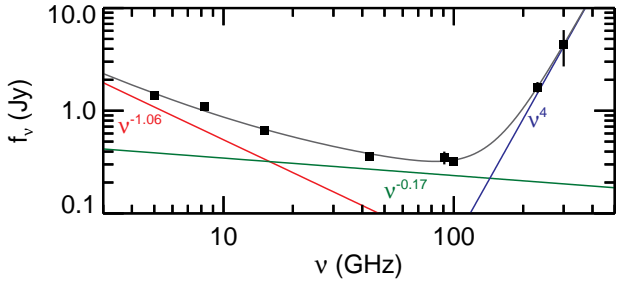}
\caption[]{The spectral energy distribution of the central 30~arcsec of NGC~253 based on the data from \citet{rodriguezrico06}.  The dark grey lines shows the function fit to the data, which comprises a synchrotron function with a power law index that was allowed to vary (shown in red), a free-free function that scales as $\nu^{-0.17}$ (shown in green), and a modified Rayleigh-Jeans function for dust that varies as $\nu^4$ (shown in blue).}
\label{f_sed}
\end{figure}

For determining the relative contribution of free-free emission to the 99.02~GHz continuum, the frequency coverage of the ALMA data by itself is insufficient for modelling the spectral energy distribution.  \citet{peel11} indicate that most of the globally-integrated 99.02~GHz emission would be continuum.  However, \citet{rodriguezrico06} found a much smaller contribution of free-free emission to the central 30~arcsec, but they assumed that the slope of the synchrotron emission is fixed by the 5 and 15~GHz data points.  If we instead repeat this fit but allow the slope of the synchrotron component to vary (as shown in Figure~\ref{f_sed}, we find that $70\pm10$\% of the 99.02~GHz emission is from free-free emission.  We applied this correction to the measured continuum emission in the rest of our analysis.

The ratio of the H40$\alpha$ flux density $f_\nu(line)$ integrated over velocity $v$ to the free-free flux density $f_\nu(cont)$ can be written as 
\begin{equation}
\begin{split}
\frac{\int f_\nu(line) dv}{f_\nu(cont)} 
  \left[\frac{\mbox{Jy}}{\mbox{Jy km s}^{-1}}\right] =
  \ \ \ \ \ \ \ \ \ \ \ \ \ \ \ \ \ \ \ \ \ \ \ \ \ \ \ \ \ \ \ \ \ \ \ \ \ \ \ \ \ \ \ \ \ \ \ \\
  5.06\times10^{32}
  \left[\frac{\epsilon_\nu}{\mbox{erg s}^{-1}\mbox{ cm}^{-3}}\right]
  \left[\frac{\nu}{\mbox{~GHz}}\right]^{-0.83}
  \left[\frac{T_e}{\mbox{~K}}\right]^{0.5}
\label{e_lcratio}
\end{split}
\end{equation}
based on equations from \citet{scoville13}.  The emissivity $\epsilon_\nu$ varies slowly as a function of electron density $n_e$ but varies strongly with $T_e$.  As shown in Table~\ref{t_spec}, we measure line-to-continuum ratios of 42.5 in the central region and $\sim$50 in the east and west regions.  We used these ratios and interpolated between the $\epsilon_\nu$ values from \citet{storey95} for case B recombination and $n_e$=$10^3$ cm$^{-3}$ to calculate $T_e$ for each of the regions in our analysis.  The $T_e$ values of 3700-4500~K we measure are slightly lower than the previously-reported values from \citep{puxley97} and \citet{rodriguezrico06}, but only by $\ltsim$1$\sigma$.  The $T_e$ values are also comparable with values measured by \citet{shaver83} and \citet{paladini04} within the central 8~kpc of the Milky Way.  If we did not apply the correction of 0.70 to the continuum emission, the measured $T_e$ would increase to values of 5800-7100~K, which are still consistent with Milky Way values.  If 50\% of the continuum emission is free-free, then the resulting $T_e$ fall within the implausible range of 2500-3000~K.

The SFR that we calculate depends upon both $T_e$ and the star formation history.  The presence of both photoionized gas and supernovae \citep[e.g.][]{lenc06, rampadarath14} imply that the star formation is continuous and has been ongoing for $>$5~Myr.  We used version 7.0.1 of the {\sc Starburst99} models \citep{leitherer99} to determine how to convert the number of ionizing photons produced as a function of time ($Q$) to SFR.  This version of Starburst99 incorporates versions of the Geneva stellar evolution tracks that incorporate rotation, with the rotation velocities of zero rotation and 40\% of the break-up velocity representing two potential extremes \citep{leitherer14}.  Using the average of the results from these two versions of the Geneva tracks for $Z$=0.040 metallicity (the highest metallicity for which tracks are available) and a \citet{kroupa02} initial mass function for a mass range of 0.1-100~M$_\odot$, we find that SFR~=~1~M$_\odot$ yr$^{-1}$ corresponds to $Q$~=~$1.85\times10^{53}$~s$^{-1}$ after $\sim$5~Myr of continuous star formation\footnotetext{http://www.cv.nrao.edu/php/splat/}\footnote{The $Q$ from the two rotation scenarios differ by a factor of $\sim$1.8.  Using $Z$=0.014 metallicity (the next lower metallicity for which Geneva tracks are available) had a $<$5\% effect on the results.  Using older stellar population models that did not include stellar rotation, \citet{kennicutt98} reported a conversion factor that would increase SFR by a factor of $\sim$2, and \citet{murphy11} gave a conversion factor that would increase SFR by 35\%.}.

Given this, SFR can be calculated from either the free-free or recombination line emission using
\begin{equation}
\begin{split}
\frac{\mbox{SFR}(line)}{\mbox{M}_\odot\mbox{ yr}^{-1}}=
  2.16\times10^{-23}
  \left[\frac{\alpha_B \mbox{ cm}^6}{\epsilon_\nu\mbox{ erg}}\right]
  \ \ \ \ \ \ \ \ \ \ \ \ \ \ \ \ \ \ \ \ \ \ \ \ \ \ \ \ \ \ \ \ \\
  \left[\frac{\nu}{\mbox{~GHz}}\right]
  \left[\frac{D}{\mbox{ Mpc}}\right]^{2}
  \left[\frac{\int f_\nu(line) dv}{\mbox{Jy km s}^{-1}}\right]
\end{split}
\end{equation}
and
\begin{equation}
\begin{split}
\frac{\mbox{SFR}(cont)}{\mbox{M}_\odot\mbox{ yr}^{-1}}=
  1.09\times10^{10}\left[\frac{\alpha_B}{\mbox{ cm}^3\mbox{ s}^{-1}}\right]
  \ \ \ \ \ \ \ \ \ \ \ \ \ \ \ \ \ \ \ \ \ \ \ \ \ \ \ \ \ \ \ \ \ \\
  \left[\frac{\nu}{\mbox{~GHz}}\right]^{0.17}
  \left[\frac{T_e}{\mbox{~K}}\right]^{0.5}
  \left[\frac{D}{\mbox{ Mpc}}\right]^{2}
  \left[\frac{f_\nu(cont)}{\mbox{Jy}}\right],
\end{split}
\end{equation}
which are adapted from \citet{scoville13}. The $\alpha_B$ term, which is the effective recombination coefficient listed by \citet{storey95}, varies by a factor of $\sim$3 between 5000 and 15000~K but negligibly with electron density within the range $10^2$-$10^5$~cm$^{-3}$.  The SFRs listed in Table~\ref{t_spec} are based on $\alpha_B$ calculated using $n_e=10^3$~cm$^{-3}$.  The two SFRs, which are mathematically linked through Equation~\ref{e_lcratio}, are limited by both the photometric accuracy and the assumptions behind the conversion between $Q$ and SFR.  Because we are adjusting $T_e$ based on the line-to-continuum ratio, the calculated SFR from the continua vary by $<$1$\sigma$ if we do not adjust the continuum emission to account for emission sources other than free-free emission.

A comparison of these results to SFRs from ultraviolet, optical, or near-infrared data would not be worthwhile, as the nuclear region is heavily obscured at those wavelengths (as discussed below).  The mid- and far-infrared continuum emission tends to be either poorly resolved or saturated in most existing data, making it impossible to calculate star formation rates from dust emission on spatial scales comparable to the ALMA data.  However, we can compare our total SFR (and $Q$, which is $(3.2\pm0.2)\times10^{53}$~s$^{-1}$ within the central 20$\times$10~arcsec) to equivalent measurements that have been made using millimetre and radio data (after rescaling all SFR and $Q$ to correspond to distances of 3.44~Mpc).  

\citet{puxley97} obtained a distance-adjusted $Q$ of $(7.0\pm1.5)\times10^{53}$~s$^{-1}$ from H40$\alpha$ data that is significantly higher than our $Q$ or other measurements of $Q$, but the line is detected at the $\sim$5$\sigma$ level, which may indicate issues with the reliability of their data.  \citet{rodriguezrico06} and \citet{kepley11} report $Q$ based on lower-frequency recombination line data that are $\sim$3$\times$ lower than our measurement.  his could be related to sensitivity issues since they had difficulty detecting line emission from all three sources that we detected.  However, these groups may be probing primarily lower-density gas in the lower-frequency data, which would lead to lower estimates of $Q$ than what we obtain with H40$\alpha$. \citet{rampadarath14}, in the latest 2.3~GHz analysis of supernova remnants in NGC~253, measure a SFR upper limit of 4.9~M$_\odot$ yr$^{-1}$; our SFR falls below this limit.  \citet{ott05} measured a distance-adjusted SFR of $4.9\pm0.5$~M$_\odot$ yr$^{-1}$ from 25~GHz continuum data which is $\sim$3$\times$ higher than our value, but they rely upon a formula originating from \citet{condon92} that uses a conversion of synchrotron emission to SFR may need to be recalibrated.  \citet{rodriguezrico06} also produced an estimate of $Q$ based on the continuum emission that is $\sim$3$\times$ lower than our value, but as we indicated above, they may have biased the free-free emission estimates to low values.

The optical recombination line emission from the centre of NGC~253 is heavily obscured, but \citet{engelbracht98} measured multiple near-infrared recombination lines, including Pa$\beta$ and Br$\gamma$, that can be compared to H40$\alpha$ emission to infer the dust attenuation to the nucleus.  If we assume that the dust functions like an attenuating sheet, then the dust attenuation $A_\lambda$ at a given infrared wavelength $\lambda$ is related to the ratio of the infrared line flux $f(\lambda)$ to the H40$\alpha$ line flux $f(H40\alpha)$ by
\begin{equation}
A_{\lambda} =
  -1.086 \ln \left\lvert
  \left[\frac{\epsilon_{H40\alpha}}{\epsilon_{\lambda}}\right]
  \left[\frac{3027~\mu\mbox{m}}{\lambda}\right]
  \left[\frac{f(\lambda)}{f(H40\alpha)}\right]
  \right\rvert .
\end{equation}
The \citet{engelbracht98} spectroscopy data were acquired within a 12$\times$2.4~arcsec aperture that also contains $\sim$70\% of the H40$\alpha$ emission in our data, so we compare the Pa$\beta$ and Br$\gamma$ fluxes from the \citet{engelbracht98} spectroscopy data (which were the lines detected at the highest signal-to-noise levels) to 0.7$\times$ our total H40$\alpha$ flux.  Assuming that $n_e=10^3$ cm$^{-3}$ and using $T_e=6200\pm400$~K from Table~\ref{t_spec}, we obtain $A_{Pa\beta}=5.0\pm0.2$ and $A_{Br\gamma}=4.2\pm0.2$.  This is higher than the values of $A_J=2.00 \pm 0.36$ and $A_K=0.87\pm0.16$ for an attenuation sheet as reported by \citet{engelbracht98}.  The results are affected by the assumed $n_e$, although the effects are on the order of 0.2 dex for a range of $10^2$-$10^5$ cm$^{-3}$.

It is likely that the attenuation was underestimated by \citet{engelbracht98}.  In their Br$\gamma$ image, they do not detect the same central peak that we detect in H40$\alpha$ (although this may be attributable to a coarser spatial resolution in the Br$\gamma$ data), and they measure a lower velocity dispersion than we do, implying that the emission from the central region is heavily obscured in the infrared data.  \citet{engelbracht98} indicate that the dust attenuation may be better described by a clumpy medium, which could affect the attenuation results.  While it is beyond the scope of this paper to explore a more complex treatment of the attenuating medium, the comparison of $A_\lambda$ values based on the attenuating sheet scenario still suggests that the attenuation measured by \citet{engelbracht98} is too low by $\sim$3 dex.

The broader implications of these results are that near-infrared recombination line emission is heavily obscured in very dusty galaxies, including starbursts like NGC~253 as well as luminous and ultraluminous infrared galaxies (galaxies with infrared luminosities $>$$10^{11}$~L$_\odot$).  SFRs based on near-infrared data may be underestimated in such objects.  Near-infrared recombination line emission has been used to calibrate some SFR calculations based on combining ultraviolet/optical star formation tracers with mid-infrared star formation tracers \citep[e.g.][]{calzetti07}.  While these formulae are probably still accurate for many galaxies, these equations may need to be recalibrated for application to more dusty objects.

\section{Summary and conclusions}

Using ALMA observations of free-free and H40$\alpha$ emission at 99.02~GHz from the centre of NGC~253, we obtain the following key results:

$\bullet$~We measure $T_e$ of 3700-4500~K within the detected regions in the centre of the galaxy, which matches previous measurements within NGC~253 as is consistent with the range of values seen in the inner 8~kpc of the Milky Way.

$\bullet$~Using both the continuum and line emission, we measure a $Q$ of $(3.2\pm0.2)\times10^{53}$~s$^{-1}$ and a SFR of $1.73\pm0.12$~M$_\odot$~yr$^{-1}$ for the central 20$\times$10~arcsec.  This measurement falls within the range of previously-published millimetre and radio results, which are inconsistent with each other, although our recombination line emission is detected at superior sensitivity levels, and our constraints on the relative contribution of free-free emission at millimetre wavelengths are more reliable.

$\bullet$~The ratio of H40$\alpha$ emission to the near-infrared line emission reported by \citet{engelbracht98} implies that the central region is obscured by $\sim$3 dex more than originally inferred.  Given this, near-infrared line emission should be used with caution to measure SFR in very dusty regions or to calibrate other SFR metrics.

This analysis represents a sample of what can potentially be achieved using ALMA to measure $T_e$ and SFR.  While this analysis is based on a target where millimetre and radio recombination line emission has been detected before, it is also based on data from ALMA Cycle 0 in which approximately one-third of the ALMA antennas were operational.  Data from dedicated observations using ALMA when it is fully operational can potentially uncover millimetre recombination line emission in more galaxies, allowing us to probe star formation in sources where other data are not usable or to calibrate other SFR metrics.

\section*{Acknowledgements}

We thank the reviewer for the useful comments on this paper.  GJB, RJB, GAF, and TWMB acknowledge support from STFC Grant ST/M000982/1.  CD acknowledges support from STFC Grant ST/L000768/1 and ERC Starting (Consolidator) Grant no.~307209.  This paper makes use of the following ALMA data: ADS/JAO.ALMA\#2011.0.001720.S.  ALMA is a partnership of ESO (representing its member states), NSF (USA) and NINS (Japan), together with NRC (Canada) and NSC and ASIAA (Taiwan), in cooperation with the Republic of Chile. The Joint ALMA Observatory is operated by ESO, AUI/NRAO and NAOJ.

{}

\label{lastpage}

\end{document}